*Article*

# Transcriptional landscape of SARS-CoV-2 infection dismantles pathogenic pathways activated by the virus, proposes unique sex-specific differences and predicts tailored therapeutic strategies

**Paolo Fagone[1], Rosella Ciurleo[2], Salvo Danilo Lombardo[3], Carmelo Iacobello[4], Concetta Ilenia Palermo[1], Yehuda Shoenfeld[5,6], Klaus Bendtzen[7], Placido Bramanti[2], Ferdinando Nicoletti[1]**

[1]Department of Biomedical and Biotechnological Sciences, University of Catania, 95123 Catania, Italy; paolofagone@yahoo.it (P.F.); ferdinic@unict.it (F.N.); ileniapalermo@libero.it (C.I.P.)
[2]IRCCS Centro Neurolesi Bonino Pulejo, C.da Casazza, 98124 Messina, Italy; rossella.ciurleo@irccsme.it; placido.bramanti@irccsme.it (P.B.); rossella.ciurleo@irccsme.it (R.C.)
[3]CeMM Research Center for Molecular Medicine of the Austrian Academy of Sciences, Lazarettgasse 14, AKH BT 25.3, A-1090 Vienna, Austria; SLombardo@cemm.oeaw.ac.at (S.D.L.)
[4]UOC Malattie Infettive, AO Cannizzaro, Catania, Italy; Carmelo.iacobello@gmail.com (C.I.)
[5]Zabludowicz Center for Autoimmune Diseases, Sheba Medical Center, affiliated to Tel-Aviv University, Israel; shoenfel@post.tau.ac.il (Y.S.)
[6]I.M. Sechenov First Moscow State Medical University of the Ministry of Health of the Russian Federation (Sechenov University); shoenfel@post.tau.ac.il (Y.S.)
[7]Rigshospitalet University Hospital, Copenhagen, Denmark; klausben@me.com (K.B.)

* Correspondence: ferdinic@unict.it (F.N.)

**Abstract:** The emergence of the severe acute respiratory syndrome coronavirus 2 (SARS-CoV-2) disease (COVID-19) has posed a serious threat to global health. As no specific therapeutics are yet available to control disease evolution, more in-depth understanding of the pathogenic mechanisms induced by SARS-CoV-2 will help to characterize new targets for the management of COVID-19. The present study identified a specific set of biological pathways altered in primary human lung epithelium upon SARS-CoV-2 infection, and a comparison with SARS-CoV from the 2003 pandemic was studied. The transcriptomic profiles were also exploited as possible novel therapeutic targets, and anti-signature perturbation analysis predicted potential drugs to control disease progression. Among them, Mitogen-activated protein kinase kinase (MEK), serine-threonine kinase (AKT), mammalian target of rapamycin (mTOR) and I kappa B Kinase (IKK) inhibitors emerged as candidate drugs. Finally, sex-specific differences that may underlie the higher COVID-19 mortality in men are proposed.

## 1. Introduction

Severe acute respiratory syndrome coronavirus-2 (SARS-CoV-2) is a member of the Coronaviridae family, isolated at the end of 2019. It was on December the 31st, 2019, that the Chinese authorities for the first time reported to the World Health Organization (WHO) a series of pneumonia cases of unknown aetiology in Wuhan City, Hubei province, China [1], and it was on January the 9th, 2020, that the Chinese Centre for Disease Prevention and Control declared that a novel coronavirus (initially named 2019-nCoV) was the causative agent [1]. The infection spreads rapidly in the population, mainly through respiratory droplets and close contact, with an incubation period ranging from 2 to 14 days [2]. The rapid propagation of the disease led the WHO to declare the state of "public health emergency of international concern" on January the 30th, 2020, and on March the 11th, 2020, the WHO declared the Pandemic [3]. As of 09/04/2020, more than 1.4 million infected and almost 90 thousand of deaths have been reported (https://www.who.int/emergencies/diseases/novel-coronavirus-2019/situation-reports/). However, the number of the infected are probably under-estimated, since the majority of cases are asymptomatic or show mild symptoms, such as dry cough, sore throat, and fever [4]. Even if there are many similarities with other coronaviruses, its high diffusion over the population in combination with the possibility to develop various fatal complications, including organ failure, septic shock, pulmonary edema, severe pneumonia, and Acute Respiratory Distress Syndrome (ARDS), make this virus a major problem for the public health [5]. It has been shown that males are more susceptible to COVID-19, reporting a prevalence between 55% and 68% [6] and increased clinical severity and mortality [7]. However, the reason behind this epidemiological difference is still not clear. In the last two months, due to the state of emergency, the scientific community has prioritized the epidemiological studies on COVID-19 infection with consequential emerging of limited information regarding the molecular basis of the disease. In this paper, we have studied the transcriptomic profile of primary human lung epithelium infected by SARS-CoV-2, focusing on the most relevant pathways modulated during the infection and correlated their role to sex genes, providing a molecular hypothesis of the gender-differences observed from the clinical data. Then, we have performed a computational analysis to find new drugs candidates, based on their ability to modulate oppositely the transcriptional profiles.

## 2. Materials and Methods

### *2.1. Dataset selection*

The NCBI Gene Expression Omnibus (GEO) database (http://www.ncbi.nlm.nih.gov/geo/) was manually searched using the MeSH term (Medical Subject Headings) "SARS-CoV-2", "COVID-19" and "SARS". The datasets were selected if they met the following criteria: (a) whole-genome transcriptomic profiling; (b) species of origin "Homo sapiens"; and (c) were not generated on cancer cell lines. Finally, two datasets were included: GSE147507 [8] and GSE47963 [9]. Briefly, the GSE147507 dataset included 3 independent biological replicates of primary human lung epithelium that were mock treated or infected with SARS-CoV-2 (USA-WA1/2020) at a MOI (Multiplicity Of Infection) of 2, for 24h. For the generation of this dataset, mRNA enriched libraries were prepared from total RNA using tTruSeq Stranded mRNA LP. cDNA libraries were sequenced using an Illumina NextSeq 500 platform. Raw sequencing reads were then aligned to the human genome (hg19) using the RNA-Seq Alignment App on Basespace (Illumina, CA, USA). The GSE47963 dataset included data from human airway epithelium cultures infected with SARS-CoV at a MOI of 2 or mock controls, for increasing time points (24-96h). Transcriptomic profiling was performed using the Agilent 4×44K human HG arrays. The submitter-supplied pre-preprocessed and normalized gene expression matrix was used for the analysis.

### *2.2. Network analysis*

The GeneMania database [10] was used to construct the network of the Differentially Expressed Genes (DEGs). Interaction data were defined as physical interaction, co-expression, predicted, co-localization, pathway, genetic interactions, and shared protein domains. The Cytoscape software [11] was used for the visualization of the network, with color-coded nodes based on the fold-change, and to perform network analysis, using the NetworkAnalyzer utility. Topological analysis was performed considering the network as undirected (i.e., containing only undirected edges). Centrality of each node in the network was ranked based on degree centrality, that corresponds to the number of edges linked to each given node. Closeness centrality and Betweeness centrality were also computed. Closeness centrality is a measure of how fast information spreads from a given node to other nodes in the network, while Betweenness centrality quantifies the number of times a node acts as a bridge along the shortest path between two other nodes. MCODE (Molecular Complex Detection) was used for module analysis of the network in Cytoscape [12]. The criteria of MCODE were as follows: degree cutoff = 2, node cutoff = 0.2, maximum depth = 100, and k-score = 2.

### *2.3. Enrichment analysis*

Functional enrichment analysis was conducted using the web-based utility, Metascape [13]. Metascape analysis is based on publicly available databases, e.g. Gene Ontology, KEGG, and MSigDB. Metascape automatically aggregates enriched terms into non-redundant groups, by calculating the pairwise similarity between any two terms [13]. Metascape uses the hypergeometric test and Benjamini–Hochberg p value correction algorithm to identify statistically significant enriched ontology terms.

For the identification of transcription factors and the comparative analysis of SARS-CoV2 induced-phenotype with the normal lung tissue, the Enrichr (http://amp.pharm.mssm.edu/Enrichr) web-based utility was used [14]. To this aim, the Encode_CHEA_Consensus_TFs and the GTEx libraries were considered. EnrichR computes the p value using the Fisher's exact test. The adjusted p-value is calculated using the Benjamini-Hochberg method for correction for multiple hypotheses testing. The z-score is computed using a modification to the Fisher exact test and assesses the deviation from the expected rank. Finally, the Combined Score is calculated by p value and the z-score (Combined Score = ln(p value) × z-score).

### *2.4. Drug Prediction Analysis*

The L1000FDW web-based utility [15] was used to identify potential drugs for the treatment of COVID-19. L1000FWD computes the similarity between a gene expression profile and the Library of Integrated Network-based Cellular Signatures (LINCS)-L1000 data, in order to prioritize drugs potentially able to reverse the input transcriptional feature [15]. The L1000 transcriptomic database belongs to the Library of Integrated Network-based Cellular Signatures (LINCS) project, a NIH Common Fund program, and includes the transcriptional profiles of ~50 human cell lines upon exposure to approximately 20,000 compounds, in a range of concentrations and time [15].

### *2.5. Statistical Analysis*

For the differential expression analysis of the GSE147507 dataset, the VOOM (mean-variance modelling at the observational level) algorithm was used. The VOOM method estimates the mean-variance relationship of the log-counts, generates a precision weight for each observation and enters these into the limma empirical Bayes analysis [16]. The analysis of the GSE47963 dataset was performed using the LIMMA function. The cloud-based application WebMeV (Multiple Experiment Viewer) was used for the statistical analyses [17]. Genes with an adjusted $p$ value $< 0.05$ and a $|\text{fold change}| > 2$ were identified as DEGs (Differentially Expressed Genes) and selected for further analysis.

Linear regression and Spearman's correlation were performed to compare the fold change of genes modulated upon SARS-CoV-2 infection and following SARS-CoV infection, at different time points.

Differences in the Combined Score for the enrichment of the lung tissue profile between women and men, stratified by age, was performed using the Mann-Whitney U test, followed by Benjamini–Hochberg multiple test correction procedure.

The GraphPad Prism (v. 8) software (San Diego, CA, USA) was used for the statistical analysis and the generation of the graphs. Unless otherwise stated, a p value<0.05 was considered for statistical significance.

## 3. Results

### *3.1. Network and enrichment analysis of SARS-CoV-2 infection*

In order to identify a specific gene signature characterizing SARS-CoV-2 infection, we first interrogated the GSE147507 dataset. We identified 129 DEGs, 94 upregulated and 35 downregulated (Figure 1A). MCODE analysis identified 7 main clusters of associated genes (Figure 1B; suppl. File 1). Gene term enrichment analysis for the upregulated genes identified several altered pathways upon SARS-CoV-2 infection, with the top three being: "cytokine-mediated signaling pathway", "IL-17 signaling pathway", and "defense response to other organism" (Fig. 1C). No significant enriched term was instead found for the downregulated DEGs. Among the statistically significant enriched terms, intracellular pathways related to NFkB, toll-like receptors and MAPK were also found (Fig. 1C). Accordingly, analysis of the transcription factors putatively involved in the regulation of the upregulated DEGs identified RELA (adj. p value=0.047), for its role to transcribe 9 out of the 94 DEGs, i.e. *BST2*, *IL32*, *TNIP1*, *ICAM2*, *TNFAIP3*, *MMP9*, *BIRC3*, *RND1* and *ICAM1* (Figure 1D). Interestingly, a number of DEGs were found to be modulated by sexual hormones, as ESR1 (Estrogen Receptor 1) was found to be involved in the regulation of 4 DEGs (*C3* and *EDN1*, among the upregulated genes; and *PDK4* and *VTCN1*, among the downregulated DEGs) and AR (Androgen Receptor) was found to regulate 6 DEGs (*CCL20* and *CXCL1*, among the upregulated genes, and *THBD*, *HEY2*, *BBOX1* and *MYLK*, among the downregulated genes) (Figure 1D and Suppl. File 1). Network analysis identified as the top hub genes, *MX1*, *IL8* and *IFITM1* (table 1).

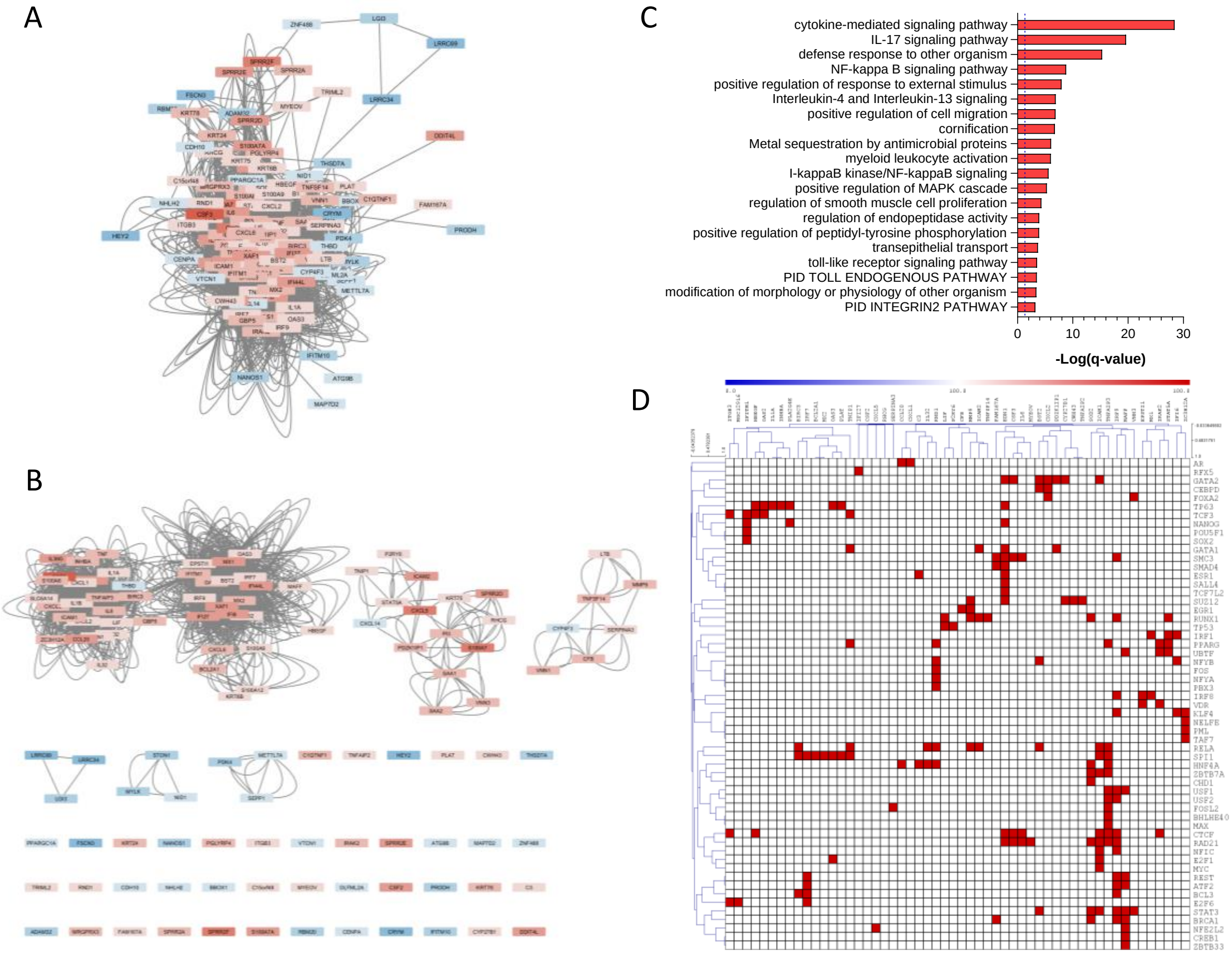

**Figure 1.** A) Gene network constructed using the Differentially Expressed Genes (DEGs) identified in the GSE147507 dataset. Nodes are color-coded based on the fold-change; B) MCODE clustering for the identification of neighborhoods where genes are densely connected; C) Gene Term enrichment using the upregulated DEGs identified in the GSE147507 dataset; D) Maps showing the potential transcription factors regulating the expression of the upregulated genes in the GSE147507 dataset

**Table 1.** Network analysis with the top 50 genes ranked based on the degree of distribution

| Gene | Degree | Betweenness Centrality | Closeness Centrality |
|---|---|---|---|
| *MX1* | 156 | 0.005556 | 0.4875 |
| *IL8* | 149 | 0.023353 | 0.573529 |
| *IFITM1* | 132 | 0.027499 | 0.52 |
| *IFI44L* | 129 | 0.001489 | 0.45 |
| *CXCL1* | 129 | 0.027552 | 0.573529 |
| *CXCL2* | 119 | 0.02143 | 0.567961 |
| *S100A8* | 118 | 0.026276 | 0.551887 |
| *IRF7* | 117 | 0.00215 | 0.466135 |
| *IFI27* | 116 | 0.022974 | 0.522321 |
| *OAS1* | 116 | 0.004078 | 0.464286 |
| *IL1B* | 116 | 0.026433 | 0.557143 |
| *IL6* | 115 | 0.007298 | 0.531818 |
| *TNFAIP3* | 112 | 0.029273 | 0.559809 |
| *ICAM1* | 112 | 0.035752 | 0.579208 |
| *XAF1* | 110 | 0.004523 | 0.483471 |
| *MX2* | 110 | 0.007159 | 0.473684 |
| *CXCL3* | 109 | 0.027718 | 0.554502 |
| *IRF9* | 106 | 0.011689 | 0.46063 |
| *OAS2* | 104 | 0.010199 | 0.481481 |
| *BST2* | 98 | 0.003963 | 0.483471 |
| *IFI6* | 91 | 0.008014 | 0.485477 |
| *S100A9* | 89 | 0.031773 | 0.541667 |
| *BIRC3* | 89 | 0.028051 | 0.549296 |
| *CCL20* | 80 | 0.027608 | 0.544186 |
| *CXCL6* | 79 | 0.016894 | 0.534247 |
| *OAS3* | 69 | 0.008639 | 0.46063 |
| *S100A12* | 62 | 0.020722 | 0.524664 |
| *PI3* | 61 | 0.035629 | 0.546729 |
| *KRT6B* | 60 | 0.036157 | 0.524664 |
| *TNFAIP2* | 59 | 0.007819 | 0.513158 |
| *TNF* | 59 | 0.030233 | 0.52 |
| *BCL2A1* | 58 | 0.007818 | 0.510917 |
| *CXCL5* | 57 | 0.011576 | 0.517699 |
| *S100A7* | 56 | 0.025749 | 0.513158 |
| *MMP9* | 56 | 0.026879 | 0.534247 |
| *SAA1* | 52 | 0.016882 | 0.524664 |
| *HBEGF* | 49 | 0.020569 | 0.4875 |
| *EPSTI1* | 48 | 0.000839 | 0.433333 |
| *CSF3* | 47 | 0.00449 | 0.50431 |
| *MAFF* | 42 | 0.019563 | 0.515419 |
| *CFB* | 42 | 0.010391 | 0.5 |
| *PDZK1IP1* | 40 | 0.018457 | 0.524664 |
| *IL32* | 40 | 0.005467 | 0.483471 |
| *LIF* | 40 | 0.006298 | 0.491597 |
| *C3* | 39 | 0.013371 | 0.50431 |
| *IL1A* | 35 | 0.008345 | 0.502146 |
| *INHBA* | 34 | 0.022095 | 0.506494 |
| *ICAM2* | 34 | 0.004401 | 0.483471 |
| *STAT5A* | 33 | 0.010319 | 0.497872 |

### *3.2. Comparison between SARS-COV-2 and SARS-CoV infection*

We next wanted to compare the gene signature induced by SARS-COV-2 and SARS-CoV infection. To this aim, we have interrogated the publicly available GSE47963 microarray dataset. We have first performed a correlation analysis on the modulation of the genes perturbed upon SARS-CoV- infection and the corresponding genes in GSE47963. To account for the differences in the experimental setting, as well as the different technologies involved, we considered all the genes with a raw p value<0.05, irrespective of the fold-change. A total of 2871 genes were found to be modulated by SARS-CoV-2 and in common with the GSE47963 dataset. As shown in Figure 3A-B, a moderate but significant correlation is found between SARS-COV-2 and SARS-CoV infection at 24h, which increases when considering SARS-CoV infection at later time points. When a more stringent selection of the DEGs is applied (i.e., adj. p value<0.05 and |fold-change|> 2), among the upregulated genes, only 1 gene is in common between SARS-CoV-2 and SARS-CoV infections at 24h (CXCL2), 6 genes are in common between SARS-CoV-2 infection and SARS-CoV infection at 48h; 25 and 22 genes are in common between SARS-CoV-2 and SARS-CoV at 72h and 96h, respectively (Figure 2C, suppl file 2). Accordingly, similar pathways are enriched between SARS-CoV-2 and SARS-CoV infection for the 72h and 96h time points (Figure 2D). Among the downregulated genes, only 1 and 4 genes are in common between SARs-CoV-2-19 and SARS-CoV at 72h and 96h, respectively (Figure 2E, suppl file 2).

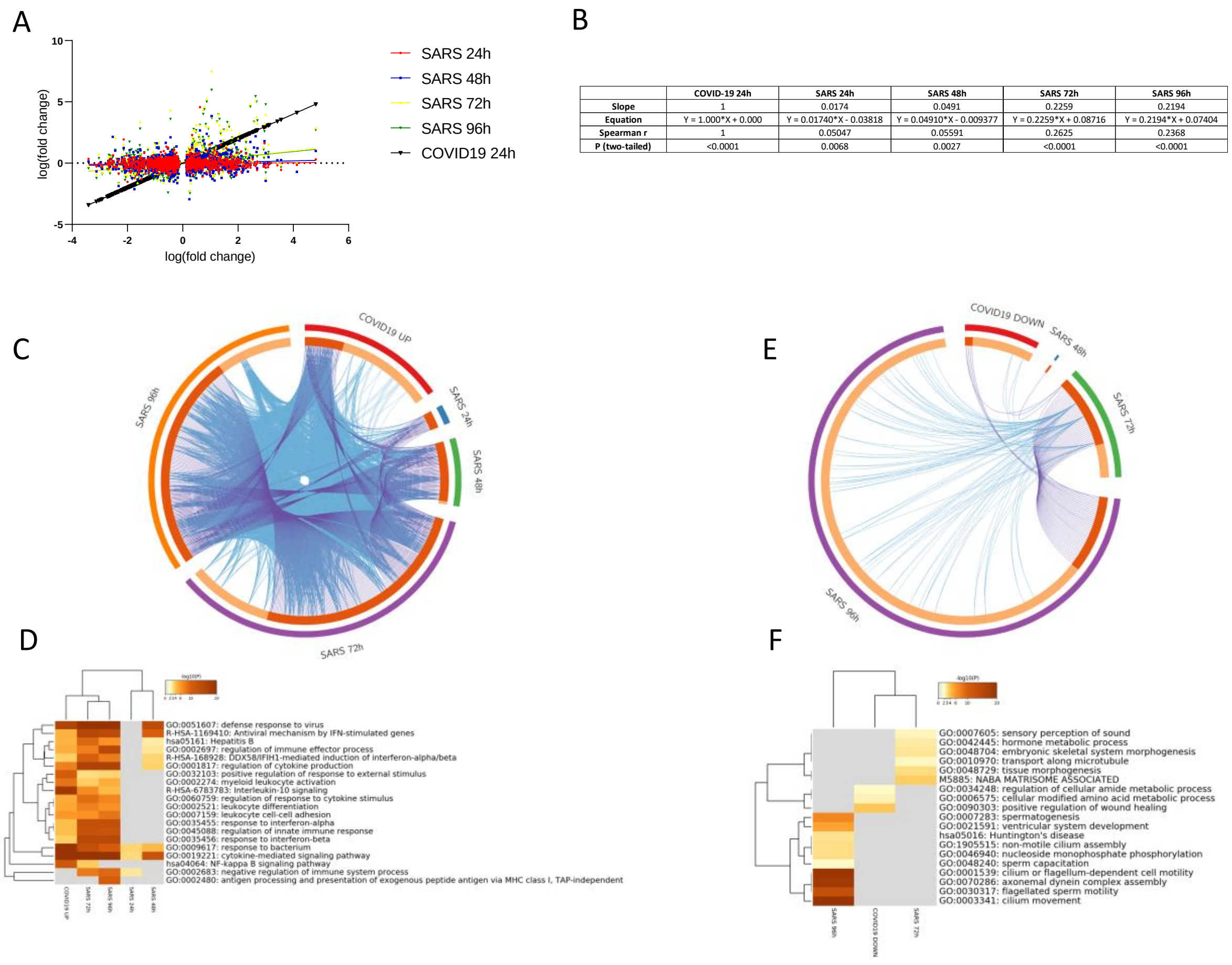

| | COVID-19 24h | SARS 24h | SARS 48h | SARS 72h | SARS 96h |
|---|---|---|---|---|---|
| Slope | 1 | 0.0174 | 0.0491 | 0.2259 | 0.2194 |
| Equation | Y = 1.000*X + 0.000 | Y = 0.01740*X - 0.03818 | Y = 0.04910*X - 0.009377 | Y = 0.2259*X + 0.08716 | Y = 0.2194*X + 0.07404 |
| Spearman r | 1 | 0.05047 | 0.05591 | 0.2625 | 0.2368 |
| P (two-tailed) | <0.0001 | 0.0068 | 0.0027 | <0.0001 | <0.0001 |

**Figure 2.** A) Scatter plot showing the correlation of gene expression between SARS-CoV-2 infection and SARS-CoV infection at different time points; B) Analysis of correlation of genes modulated upon SARS-CoV-2 and SARS-CoV infection, at different time points; C) Circos plot showing the overlapping between the genes significantly upregulated following SARS-CoV-2 infection and genes upregulated upon SARS-CoV infection at different time points. Purple lines link the same genes that are shared by the input lists. Blue lines link the different genes that fall in the same ontology term; D) Hierarchical clustering of the top most enriched terms by genes significantly upregulated upon infection. The heatmap is colored by the p values, and grey cells indicate the lack of significant enrichment; E) Circos plot showing overlapping between the genes significantly downregulated following SARS-CoV-2 infection and genes downregulated upon SAR-CoV infection at different time points. Purple lines link the same genes that are shared by the input lists. Blue lines link the different genes that fall in the same ontology term; F) Hierarchical clustering of the top most enriched terms by the downregulated genes upon infection. The heatmap is colored by the p values, and grey cells indicate the lack of significant enrichment;

### 3.3. *Drug prediction analysis*

Anti-signature perturbation analysis was performed using the DEGs identified for SARS-CoV-2 infection (Figure 3A). In Table 2, we have enlisted the potential drugs identified by the L1000FWD analysis. Among them, the top three drugs are: BRD-K23875128, a Rho kinase inhibitor; SA-792728, a sphingosine kinase inhibitor; and sirolimus, an mTOR inhibitor. Figure 3B shows the relative percentage of the drugs based on their mode of action, which identifies the inhibitors of MEK, as the most represented drug category (Figure 3B).

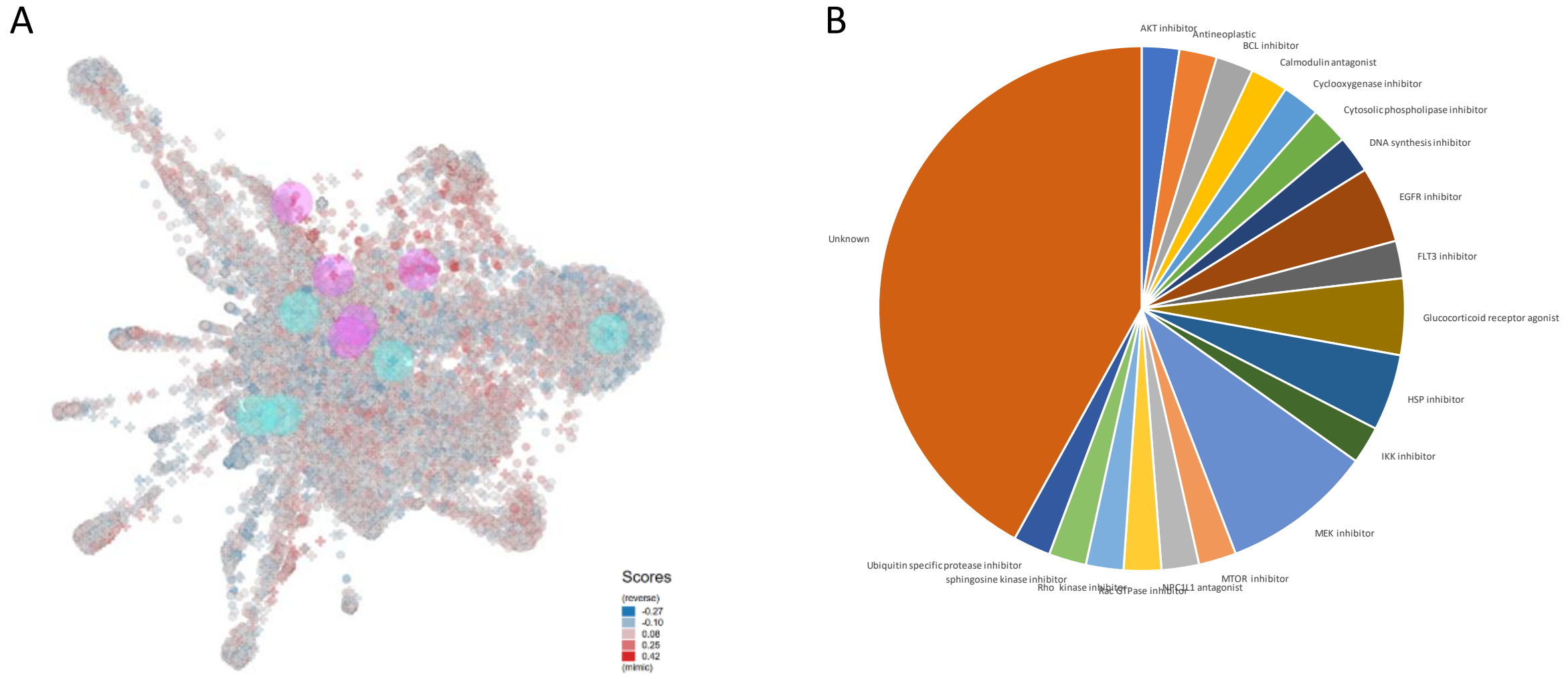

**Figure 3.** A) L1000FDW visualization of drug-induced signature. Input genes are represented by the significantly upregulated and downregulated genes obtained from the analysis of the GSE147507 dataset, Blue and red circles identify drugs with similar and anti-similar signatures. Dots are color-coded based on the similarity score; B) Percentage of drug categories identified by the L1000FDW analysis

**Table 2.** List of potential drugs for SARS-CoV-2 infection as identified by the L1000FWD analysis

| drug | similarity score | P value | Q value | Z score | combined score | mode of action |
|---|---|---|---|---|---|---|
| BRD-K23875128 | -0.2717 | 3.53E-23 | 3.08E-20 | 1.82 | -40.95 | Rho kinase inhibitor |
| SA-792728 | -0.2391 | 3.53E-18 | 1.11E-15 | 1.75 | -30.61 | sphingosine kinase inhibitor |
| sirolimus | -0.2391 | 2.10E-18 | 7.19E-16 | 1.77 | -31.33 | MTOR inhibitor |
| BRD-K44366801 | -0.2283 | 4.53E-17 | 1.17E-14 | 1.75 | -28.62 | Unknown |
| TPCA-1 | -0.2174 | 5.15E-16 | 1.00E-13 | 1.8 | -27.53 | IKK inhibitor |
| selumetinib | -0.2174 | 6.28E-16 | 1.16E-13 | 1.77 | -26.83 | MEK inhibitor |
| ezetimibe | -0.2174 | 8.57E-16 | 1.53E-13 | 1.78 | -26.83 | Niemann-Pick C1-like 1 protein antagonist\|Cholesterol inhibitor |
| desoximetasone | -0.2174 | 6.53E-16 | 1.20E-13 | 1.74 | -26.47 | Glucocorticoid receptor agonist |
| BRD-K60070073 | -0.2174 | 1.44E-16 | 3.26E-14 | 1.69 | -26.84 | Unknown |
| TPCA-1 | -0.2174 | 2.70E-16 | 5.72E-14 | 1.8 | -27.95 | IKK inhibitor |
| BRD-K03601405 | -0.2065 | 9.63E-16 | 1.69E-13 | 1.79 | -26.9 | Unknown |
| BRD-K88622704 | -0.2065 | 4.67E-15 | 7.02E-13 | 1.75 | -25.08 | Unknown |
| CT-200783 | -0.2065 | 1.73E-15 | 2.88E-13 | 1.82 | -26.92 | Unknown |
| CAM-9-027-3 | -0.2065 | 4.81E-16 | 9.49E-14 | 1.89 | -29.01 | Unknown |
| BRD-K25373946 | -0.1957 | 1.70E-14 | 2.28E-12 | 1.82 | -25.07 | Unknown |
| piperlongumine | -0.1957 | 4.30E-14 | 5.35E-12 | 1.79 | -23.92 | Unknown |
| BRD-K89687904 | -0.1957 | 3.43E-14 | 4.38E-12 | 1.79 | -24.07 | PKC inhibitor |
| NSC-632839 | -0.1957 | 9.94E-14 | 1.11E-11 | 1.65 | -21.47 | Ubiquitin specific protease inhibitor |
| BRD-K03371390 | -0.1957 | 7.20E-14 | 8.33E-12 | 1.77 | -23.28 | Unknown |
| BRD-K06765193 | -0.1957 | 3.18E-14 | 4.10E-12 | 1.73 | -23.39 | Unknown |
| BRD-K32101742 | -0.1957 | 7.46E-14 | 8.56E-12 | 1.67 | -21.95 | Unknown |
| WZ-4002 | -0.1848 | 1.25E-12 | 1.06E-10 | 1.64 | -19.48 | EGFR inhibitor |
| U-0126 | -0.1848 | 1.10E-12 | 9.46E-11 | 1.69 | -20.16 | MEK inhibitor |
| piperlongumine | -0.1848 | 2.30E-13 | 2.36E-11 | 1.84 | -23.29 | Unknown |
| radicicol | -0.1848 | 3.85E-13 | 3.78E-11 | 1.75 | -21.75 | HSP inhibitor |
| ABT-737 | -0.1848 | 3.40E-12 | 2.68E-10 | 1.73 | -19.88 | BCL inhibitor |
| arachidonyl-trifluoro-methane | -0.1848 | 2.67E-13 | 2.68E-11 | 1.86 | -23.43 | Cytosolic phospholipase inhibitor |
| fluticasone | -0.1848 | 6.79E-13 | 6.35E-11 | 1.81 | -22.07 | Glucocorticoid receptor agonist |
| BRD-K23875128 | -0.1848 | 1.53E-12 | 1.28E-10 | 1.78 | -21.02 | Rho kinase inhibitor |
| tyrphostin-AG-1296 | -0.1848 | 4.14E-13 | 4.02E-11 | 1.87 | -23.13 | FLT3 inhibitor |
| NVP-AUY922 | -0.1739 | 9.17E-12 | 6.31E-10 | 1.62 | -17.92 | HSP inhibitor |
| TPCA-1 | -0.1739 | 1.67E-11 | 1.05E-09 | 1.8 | -19.34 | IKK inhibitor |
| TPCA-1 | -0.1739 | 7.41E-13 | 6.81E-11 | 1.91 | -23.18 | IKK inhibitor |
| ST-4070169 | -0.1739 | 8.32E-12 | 5.84E-10 | 1.73 | -19.21 | Unknown |
| valdecoxib | -0.1739 | 1.67E-11 | 1.05E-09 | 1.78 | -19.18 | Cyclooxygenase inhibitor |
| EMF-bca1-16 | -0.1739 | 9.17E-12 | 6.31E-10 | 1.68 | -18.55 | Unknown |
| BIBU-1361 | -0.1739 | 7.06E-12 | 5.10E-10 | 1.82 | -20.29 | EGFR inhibitor |
| MD-II-038 | -0.1739 | 1.11E-11 | 7.37E-10 | 1.74 | -19.11 | Unknown |
| methoxsalen | -0.1739 | 1.48E-11 | 9.49E-10 | 1.79 | -19.39 | DNA synthesis inhibitor |
| MK-2206 | -0.1739 | 8.88E-12 | 6.18E-10 | 1.8 | -19.91 | AKT inhibitor |
| TPCA-1 | -0.1739 | 1.11E-11 | 7.37E-10 | 1.83 | -20.02 | IKK inhibitor |
| phenethyl-isothiocyanate | -0.1739 | 8.06E-12 | 5.69E-10 | 1.82 | -20.21 | Antineoplastic |
| BRD-A09984573 | -0.1739 | 8.32E-12 | 5.84E-10 | 1.71 | -18.98 | Unknown |
| BRD-K12244279 | -0.1739 | 9.79E-12 | 6.66E-10 | 1.69 | -18.64 | MEK inhibitor |
| PD-198306 | -0.163 | 8.94E-11 | 4.65E-09 | 1.7 | -17.04 | MEK inhibitor |
| BRD-K18726304 | -0.163 | 6.99E-11 | 3.76E-09 | 1.79 | -18.18 | Unknown |
| calmidazolium | -0.163 | 4.07E-11 | 2.36E-09 | 1.73 | -17.97 | Calmodulin antagonist |
| NSC-23766 | -0.163 | 7.44E-11 | 3.99E-09 | 1.79 | -18.15 | Rac GTPase inhibitor |
| BRD-K66037923 | -0.163 | 1.58E-10 | 7.51E-09 | 1.75 | -17.1 | Unknown |
| EMF-sumo1-11 | -0.163 | 2.87E-10 | 1.28E-08 | 1.65 | -15.78 | Unknown |

*3.4. Similarity between the SARS-CoV-2 -related phenotype and the healthy lung tissue males and females*

Analysis of the sex-specific enrichment of the COVID-19-related gene signature in the lung tissue of healthy subjects was performed using the GTex library, implemented in the EnrichR utility. Overall, 47 and 87 samples of lung tissue were tested from females and males, respectively. As shown in Figure 4, in the fourth decade of life, the female lung tissues has a more similar phenotype to that induced upon SARS-CoV-2 infection than identical tissues from men (p=0.019 by Mann-Whitney U test). No significant differences were observed between male and female tissues in other decades of age, although a similar trend is observed at the age of 50-59 years (Figure 4)

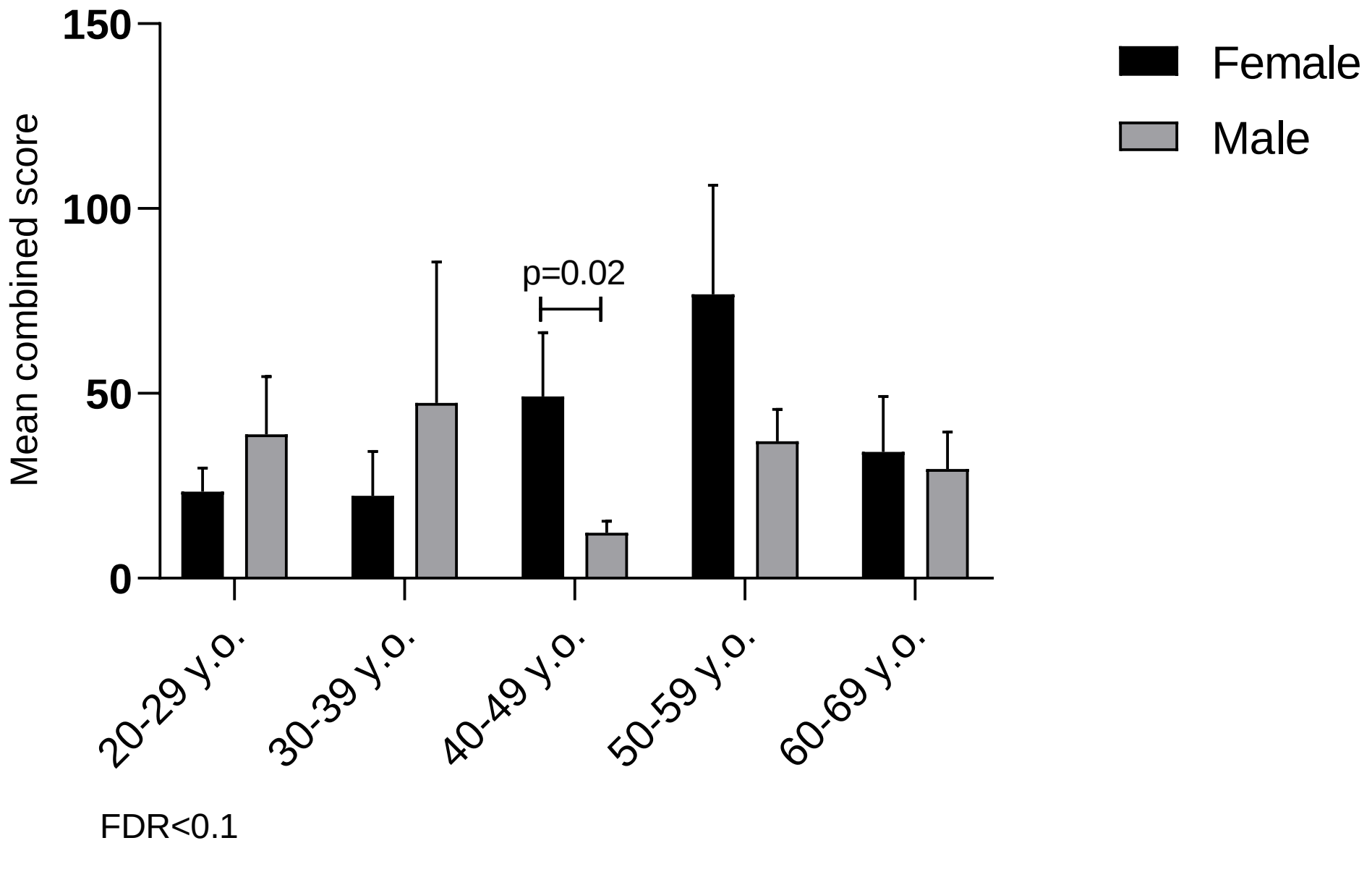

**Figure 4.** Combined score of the similarity between the SARS-CoV-2-related phenotype and the healthy lung tissue

## 4. Discussion

Even though recent evidence indicate that most cases of COVID-19 have an asymptomatic or subclinical course of infection, a significant proportion of patients develop flu-like symptoms of variable severities that may require hospitalization. In addition, a group of high risk individuals has been identified that include older people, individuals affected by other comorbidities, such as diabetes and hypertension, and those with a history of smoking, who are susceptible to develop a severe course of SARS-CoV-2 infection [18–20]. The percentage of lethality due to COVID-19 infection varies greatly between different countries, for example from ~11% in Italy to ~1% in neighboring Germany. The reasons for this are not entirely clear although some countries express lethality as case fatality rates, others as infection fatality rates; the latter depending on the SARS-CoV-2 testing capacity of individual countries. Different lethalities also depend on patient-related factors, including comorbidities especially in older individuals. Development and severity of COVID-19 also appears to depend on individual propensities for massive release of proinflammatory and immunoactivating cytokines including interleukin (IL)-1β, IL-2, IL-6, IL-7, IL-8, tumor necrosis factor-$\alpha$ (TNF-$\alpha$) and chemokines (CXC-chemokine ligand 10 (CXCL10) and CC-chemokine ligand 2 (CCL2) at the level of the infected lung tissues and accumulating immune cells [2,21] developing a reaction known as Cytokines Release Syndrome (CRS). This is likely to promote self-sustaining inflammatory processes that may contribute to the development of respiratory failure and systemic, possibly lethal, manifestations similar to those seen in patients with e.g. meningococcal sepsis: hyperthermia, catastrophic multiple organ failure, shock and disseminated intravascular coagulation [22]. It is of interest that CRS characterizes secondary haemophagocytic lymphohistiocytosis (sHLH), that is an under-recognized, hyperinflammatory syndrome characterized by a fulminant and fatal hypercytokinaemia with multiorgan failure. In adults, sHLH is most commonly triggered by viral infections and occurs in 3·7–4·3% of sepsis cases [21].

Predictors of fatality from a recent retrospective, multicenter study of 150 confirmed COVID-19 cases in Wuhan, China, included elevated ferritin (mean 1297·6 ng/ml in non-survivors vs 614·0 ng/ml in survivors; $p<0·001$) and IL-6 ($p<0·0001$), suggesting that mortality might be due to virally driven hyperinflammation [2,21].

A "cytokine storm" underlying lethal outcomes of COVID-19 is also supported by anecdotical though repeated observations that the anti-IL-6 receptor antibody, tocilizumab, appears to beneficially influence the course of COVID-19. This has led to the initiation of Phase II and Phase III studies of tocilizumab in these patients [23]. Other anti-inflammatory therapeutics, including glucocorticoids, JAK inhibitors and choloroquine/hydrocholoroquine, may also improve patients' prognosis [24]. It should be noted, however, that the ability of glucocorticoids to block production of the potentially most dangerous cytokines, TNF-$\alpha$ and IL-1$\beta$, occurs largely at the genomic level therefore requiring several hours for effective manifestation. This hampers the effectiveness of glucocorticoid therapies in catastrophic situations where high amounts of these cytokines have already been released to the circulation. In contrast, specific anti-cytokine antibodies or receptor antagonists clinical approved on other indications would be expected to have an immediate neutralizing effect that might reduce mortality.

The identification of the precise pathogenic mechanisms by which SARS-CoV-2 induces organ damage are of immediate urgency. Unfortunately, emerging data seem to indicate that other organs such as heart, kidney and the central nervous system may also be attacked, albeit in a more silent fashion, by SARS-CoV-2 [25,26]. Indeed, patients may show proteinuria, elevated baseline serum creatinine levels and hematuria [25], as well as neurological symptoms, that include headache, epilepsy, disturbed consciousness, anosmia and dysgeusia [26]. Some COVID-19 patients also develop a thrombogenic diathesis that is characterized by marked elevation of D-dimer and other procoagulant parameters diverging from those observed during SARS infections [27]. Thus, anticoagulant treatment is associated with decreased mortality in severe COVID-19 patients with coagulopathy [28]. Understanding whether or not a cytokine storm also contributes to damage to these organs or if other pathogenetic mechanisms operate is clearly needed to develop organ-tailored therapeutic approaches.

While waiting for the eventual advent of a vaccine, other drugs used for different indications may be repurposed to treat COVID-19 patients, as well. These include a few antiviral options under clinical trial, e.g. remdesivir [29], and lopinavir/ritonavir given alone or in combination with interferon-β and hydroxychloroquine alone or with azitromicin (ClinicalTrials.gov identifier: NCT04332107; NCT04339816; NCT04336332; NCT04332094; NCT04335552; NCT04322123).

The use of whole-genome expression data has been extensively used by ourselves and others for the identification of novel pathogenic pathways and therapeutic targets in several human pathologies including, autoimmune diseases [30–32] and cancer [33,34]. Network analysis allows to extensive testing of different biological features [35] and to extrapolate new information that otherwise would be lost. Here, we have used computational methods and network analysis to better characterize the cellular response to SARS-CoV-2 infection, suggesting the molecular basis for the observed gender differences and to predict possible new therapeutic targets. We exploited a publicly available RNA-seq dataset, GSE147507, generated from primary lung cells upon infection with SARS-CoV-2 thus extending analyses of the original data published by Blanco-Melo et al. [8]. We focused on comparisons of the transcriptional signature induced by SARS-CoV-2 and SARS-CoV from the 2003 pandemic and on the molecular mechanisms likely involved in gender differences in COVID-19 susceptibility. Using an in silico-pharmacology approach, we have also presently predicted potential drugs for COVID-19 treatment. Our work differs substantially from the study by Guzzi and colleagues [36], as their study explores the SARS-CoV-2/host receptor recognition and makes use of predicted functional interactions based on convergent SARS-CoV and Middle East respiratory syndrome coronavirus (MERS-CoV) transcriptional profiles generated on established lung cancer cells [36].

Our study revealed that many differentially expressed genes are involved in inflammation and response to external organisms, such as *S100A7A*, *CSF3*, *ICAM2*, *CCL20*, *CXCL3*, *IL6*, *CSF2*, *CXCL5*, *IL8*, *ICAM1*, *OAS1*. These genes are known to promote the recruitment and activation of granulocytes, monocytes and macrophages proliferation and infiltration, that, as mentioned before, is one of the key features of COVID-19. As expected, the most important transcription factor regulating the response to COVID-19 is RELA, which is involved in NF-kB formation and controls both proliferative and inflammatory cellular responses [37]. Interestingly, other representative transcription factors in our network are implicated in mechanisms of DNA damage and repairing and chromatin remodeling events, such as RAD21, CTCF, SPI1 and GATA2 [38–41].

To better understand the similarities between SARS infection, we also compared the transcriptomic profile induced by SARS-CoV-2 infection and SARS infections at different time points. The results show that the transcriptomic signature induced by SARS-CoV-2 infection correlates better with the latest stages of SARS infection and, accordingly, many biological pathways related to cytokine response and host defense mechanisms are shared between COVID-19 24h and SARS 72h and 96h. Hence, we propose that the accelerated cellular response to SARS-CoV-2, in comparison to SARS, may explain the higher ability of SARS-CoV-2 to spread [2].

Next, by using an anti-signature analysis approach [42,43], we have predicted possible novel drug options for the treatment of SARS-CoV-2 infection. Even if the mechanism of action of most of the predicted target molecules is not yet known, other targeting drugs are anti-inflammatory (glucocorticoids) and anti-proliferative drugs, such as Mitogen-activated protein kinase kinase (MEK), serine-threonine kinase (AKT), mammalian target of rapamycin (mTOR) and I kappa B Kinase (IKK) inhibitors. In agreement with data by Zhou et al. [44], our analysis shows that the mTOR inhibitor, sirolimus, may be a candidate drug for use in COVID-19 patients. Moreover, in an in vitro study, extracellular signal-regulated kinase (ERK)/mitogen-activated protein kinase (MAPK) and phosphoinositol 3-kinase (PI3K)/AKT/mTOR signaling responses were previously found to be modulated in response to the infection with another coronavirus, MERS-CoV [45]. Studies have also shown that the mammalian target of rapamycin complex 1 (mTORC1) is a key factor in regulating the replication of viruses [45,46], and in patients with H1N1 pneumonia, early treatment with corticosteroids and an rapamycin has been associated with improvement in hypoxia, multiple organ dysfunction, virus clearance, and shortened time in ventilators [47]. Low-dose inhibitors of mTORC1 are also beneficial in the elderly by increasing immune function and reducing infections and complications [48]. We have also generated in vivo proof-of-concept that sirolimus is effective in prevention of HIV replication in a humanized SCID model [49] and have proposed the use of rapamycin in HIV infection and its complications [50,51]. Subsequently, our in vivo data have been replicated in the same model by Heredia et al., using an ATPase inhibitor of mTOR [52], and it has been further extended by Latinoci et al., in a different mouse model of HIV infection, where they showed the synergistic antiretroviral action of rapamycin with standard antiretroviral therapy [53]. Recent data have also demonstrated a possible antiretroviral action of rapamycin in HIV-infected kidney transplant recipients. It appears therefore that an antiviral mode of action of rapamycin coupled with its immunomodulatory potential that may dampen excessive production of proinflammatory cytokines would warrant clinical studies with this drug in selected patients with COVID-19 [54].

Despite this, we have to acknowledge the limitations of the present study. First, the differentially expressed genes, that we have prioritized in our study, have been obtained from the analysis of an in vitro model of SARS-CoV-2 infection, hence the data may be incomplete, and functional associations occurring in patients may be missing. Studies integrating lung-specific gene expression profiles with the SARS-CoV-2 infection-related gene signature may help to better identify potential repurposable drugs. Also, although some of our findings have been confirmed by various literature data, most of the predicted repurposable drugs need to be validated and, for several of them, preclinical studies are warranted to evaluate in vivo efficiency and side effects before clinical trials. Indeed, our approach cannot identify dose–response and dose–toxicity effects for the candidate drugs. Furthermore, drug combinations targeting multiple pathogenic pathways can be envisioned, as it has been the case for standard antiretroviral therapy with rapamycin in a mouse model of HIV [53] . Despite these limitations, our study can minimize the translational gap between preclinical data and clinical application, which appears an issue of paramount relevance in view of the unprecedent social and epidemiological emergence provoked from the rapid escalation of the SARS-CoV-2 outbreak.

To decipher the reasons for the gender differences in COVID-19 susceptibility [2], we have compared the transcriptomic profile of lung tissue from healthy women and men with the transcriptomic induced by COVID-19. At ages between 40 and 60 years, the transcriptomic feature of the female lung tissue was more similar to those induced by COVID-19 than in male tissue. A lower threshold of acute response to SARS-CoV-2 infection in men may at least partly explaine the reduced incidence of COVID-19 in women. This hypothesis will need to be weighed and validated on the ground of appropriate epidemiological data that will allow to ascerain whether the apparent protection from COVID-19 infection in famales is specifically observed in the above-indicated range of age. If so, next question will be whether this ”physiological” observation may be translated into therapeutic opportunity by dismantling those potential factors that might have contributed to the induction of this ”COVID-19-resistant lung phenotype” in women of these ages. Clearly, female-specific hormonal factors for example occurring just before or during the menopausal period can be implicated.
In this regard, it is notable that among the SARS-CoV-2-induced genes encoding the neutrophil chemotactic factor CXCL1 and the predominantly dendritic cell chemotactic factor CCL20 are regulated by androgen receptor AR (androgen receptor), whereas *C3* and *EDN1* are regulated by ESR1 (estrogen receptor 1). AR plays a role in innate and adaptive immune regulations [55,56], especially in the recruitment of neutrophils and macrophages, that are been proven to be strongly associated with COVID-19 in lung tissue [24]. Both *CXCL1* and *CCL20* emerged in the top 50 ranking nodes in our network ordered by degree, and both are modulated even during SARS infection [57,58]. This suggests their involvement in different coronavirus infections and, in addition, a different role in female and male responses to these infections.
It is noteworthy that ER is involved as regulators of immune responses by enhancing interferon production and anti-viral response [59], and that selective oestrogen receptor modulators have been proposed as potential drugs to treat coronavirus infection [44]. Toremifene, for example, potentially affects several key host proteins associated with coronavirus: RPL19, HNRNPA1, NPM1, EIF3I, EIF3F, and EIF3E. Taken together, these convergent observations point to mechanisms that may explain the lower female incidence and/or lethality of COVID-19 offering candidate therapeutic options in patients with SARS-CoV-2 infection.

## 5. Conclusions

COVID-19 is a severe infection currently spreading as a pandemic. Here, we have investigated the transcriptomic profile of primary human lung cells upon infection with SARS-CoV-2, characterizing the most activated intracellular pathways and in order to provide a molecular explanation for the gender differences in the clinical manifestations. Finally, we have identified new potential drugs for COVID-19 therapies. We found that targeting the mTOR pathway could be a promising therapeutic avenue to fight COVID-19, improving symptomatology and reducing mortality rates.

**Author Contributions:** Conceptualization, Paolo Fagone, Yehuda Shoenfeld, Klaus Bendtzen and Ferdinando Nicoletti; Data curation, Paolo Fagone and Carmelo Iacobello; Formal analysis, Rosella Ciurleo, Salvo Danilo Lombardo and Concetta Ilenia Palermo; Funding acquisition, Placido Bramanti; Investigation, Salvo Danilo Lombardo; Methodology, Paolo Fagone; Project administration, Placido Bramanti; Supervision, Ferdinando Nicoletti; Writing – original draft, Rosella Ciurleo, Salvo Danilo Lombardo and Concetta Ilenia Palermo; Writing – review & editing, Carmelo Iacobello, Yehuda Shoenfeld, Klaus Bendtzen, Placido Bramanti and Ferdinando Nicoletti.

**Funding:** This study was supported by current research funds 2020 of IRCCS "Centro Neurolesi Bonino-Pulejo", Messina, Italy.

**Conflicts of Interest:** The authors declare no conflict of interest

## References

[1] Zhu N, Zhang D, Wang W, Li X, Yang B, Song J, et al. A Novel Coronavirus from Patients with Pneumonia in China, 2019. N Engl J Med 2020;382:727–33. doi:10.1056/NEJMoa2001017.

[2] Xu J, Zhao S, Teng T, Abdalla AE, Zhu W, Xie L, et al. Systematic Comparison of Two Animal-to-Human Transmitted Human Coronaviruses: SARS-CoV-2 and SARS-CoV. Viruses 2020;12. doi:10.3390/v12020244.

[3] Cucinotta D, Vanelli M. WHO Declares COVID-19 a Pandemic. Acta Biomed 2020;91:157–60. doi:10.23750/abm.v91i1.9397.

[4] Sohrabi C, Alsafi Z, O'Neill N, Khan M, Kerwan A, Al-Jabir A, et al. World Health Organization declares global emergency: A review of the 2019 novel coronavirus (COVID-19). Int J Surg 2020. doi:10.1016/j.ijsu.2020.02.034.

[5] Chen N, Zhou M, Dong X, Qu J, Gong F, Han Y, et al. Epidemiological and clinical characteristics of 99 cases of 2019 novel coronavirus pneumonia in Wuhan, China: a descriptive study. Lancet 2020. doi:10.1016/S0140-6736(20)30211-7.

[6] Cai H. Sex difference and smoking predisposition in patients with COVID-19. Lancet Respir Med 2020. doi:10.1016/s2213-2600(20)30117-x.

[7] Li L, Huang T, Wang Y, Wang Z, Liang Y, Huang T, et al. 2019 novel coronavirus patients' clinical characteristics, discharge rate and fatality rate of meta-analysis. J Med Virol 2020. doi:10.1002/jmv.25757.

[8] Blanco-Melo D, Nilsson-Payant B, Liu W-C, Moeller R, Panis M, Sachs D, et al. SARS-CoV-2 launches a unique transcriptional signature from in vitro, ex vivo, and in vivo systems. BioRxiv 2020:2020.03.24.004655. doi:10.1101/2020.03.24.004655.

[9] Mitchell HD, Eisfeld AJ, Sims AC, McDermott JE, Matzke MM, Webb-Robertson BJM, et al. A Network Integration Approach to Predict Conserved Regulators Related to Pathogenicity of Influenza and SARS-

CoV Respiratory Viruses. PLoS One 2013;8. doi:10.1371/journal.pone.0069374.

[10] Warde-Farley D, Donaldson SL, Comes O, Zuberi K, Badrawi R, Chao P, et al. The GeneMANIA prediction server: biological network integration for gene prioritization and predicting gene function. Nucleic Acids Res 2010;38:W214-20. doi:10.1093/nar/gkq537.

[11] Reimand J, Isserlin R, Voisin V, Kucera M, Tannus-Lopes C, Rostamianfar A, et al. Pathway enrichment analysis and visualization of omics data using g:Profiler, GSEA, Cytoscape and EnrichmentMap. Nat Protoc 2019;14:482–517. doi:10.1038/s41596-018-0103-9.

[12] Bader GD, Hogue CWV. An automated method for finding molecular complexes in large protein interaction networks. BMC Bioinformatics 2003;4. doi:10.1186/1471-2105-4-2.

[13] Zhou Y, Zhou B, Pache L, Chang M, Khodabakhshi AH, Tanaseichuk O, et al. Metascape provides a biologist-oriented resource for the analysis of systems-level datasets. Nat Commun 2019;10:1523. doi:10.1038/s41467-019-09234-6.

[14] Kuleshov M V, Jones MR, Rouillard AD, Fernandez NF, Duan Q, Wang Z, et al. Enrichr: a comprehensive gene set enrichment analysis web server 2016 update. Nucleic Acids Res 2016;44:W90-7. doi:10.1093/nar/gkw377.

[15] Wang Z, Lachmann A, Keenan AB, Ma'Ayan A. L1000FWD: Fireworks visualization of drug-induced transcriptomic signatures. Bioinformatics 2018;34:2150–2. doi:10.1093/bioinformatics/bty060.

[16] Law CW, Chen Y, Shi W, Smyth GK. Voom: Precision weights unlock linear model analysis tools for RNA-seq read counts. Genome Biol 2014;15. doi:10.1186/gb-2014-15-2-r29.

[17] Wang YE, Kutnetsov L, Partensky A, Farid J, Quackenbush J. WebMeV: A Cloud Platform for Analyzing and Visualizing Cancer Genomic Data. Cancer Res 2017;77:e11–4. doi:10.1158/0008-5472.CAN-17-0802.

[18] Yang J, Zheng Y, Gou X, Pu K, Chen Z, Guo Q, et al. Prevalence of comorbidities in the novel Wuhan coronavirus (COVID-19) infection: a systematic review and meta-analysis. Int J Infect Dis 2020. doi:10.1016/j.ijid.2020.03.017.

[19] Guo W, Li M, Dong Y, Zhou H, Zhang Z, Tian C, et al. Diabetes is a risk factor for the progression and prognosis of COVID-19. Diabetes Metab Res Rev 2020:e3319. doi:10.1002/dmrr.3319.

[20] Emami A, Javanmardi F, Pirbonyeh N, Akbari A. Prevalence of Underlying Diseases in Hospitalized Patients with COVID-19: a Systematic Review and Meta-Analysis. Arch Acad Emerg Med 2020;8:e35.

[21] Mehta P, McAuley DF, Brown M, Sanchez E, Tattersall RS, Manson JJ, et al. COVID-19: consider cytokine storm syndromes and immunosuppression. Lancet (London, England) 2020;395:1033–4. doi:10.1016/S0140-6736(20)30628-0.

[22] Tang N, Li D, Wang X, Sun Z. Abnormal coagulation parameters are associated with poor prognosis in patients with novel coronavirus pneumonia. J Thromb Haemost 2020;18:844–7. doi:10.1111/jth.14768.

[23] Lu C-C, Chen M-Y, Chang Y-L. Potential therapeutic agents against COVID-19: What we know so far. J Chin Med Assoc 2020. doi:10.1097/JCMA.0000000000000318.

[24] Zhang W, Zhao Y, Zhang F, Wang Q, Li T, Liu Z, et al. The use of anti-inflammatory drugs in the treatment of people with severe coronavirus disease 2019 (COVID-19): The experience of clinical immunologists from China. Clin Immunol 2020. doi:https://doi.org/10.1016/j.clim.2020.108393.

[25] Cheng Y, Luo R, Wang K, Zhang M, Wang Z, Dong L, et al. Kidney disease is associated with in-hospital death of patients with COVID-19. Kidney Int 2020. doi:10.1016/j.kint.2020.03.005.

[26] Wu Y, Xu X, Chen Z, Duan J, Hashimoto K, Yang L, et al. Nervous system involvement after infection with COVID-19 and other coronaviruses. Brain Behav Immun 2020. doi:10.1016/J.BBI.2020.03.031.

[27] Han H, Yang L, Liu R, Liu F, Wu K, Li J, et al. Prominent changes in blood coagulation of patients with SARS-CoV-2 infection. Clin Chem Lab Med 2020;0. doi:10.1515/cclm-2020-0188.

[28] Tang N, Bai H, Chen X, Gong J, Li D, Sun Z. Anticoagulant treatment is associated with decreased mortality in severe coronavirus disease 2019 patients with coagulopathy. J Thromb Haemost 2020. doi:10.1111/jth.14817.

[29] Wang M, Cao R, Zhang L, Yang X, Liu J, Xu M, et al. Remdesivir and chloroquine effectively inhibit the recently emerged novel coronavirus (2019-nCoV) in vitro. Cell Res 2020. doi:10.1038/s41422-020-0282-0.

[30] Lombardo SD, Mazzon E, Basile MS, Cavalli E, Bramanti P, Nania R, et al. Upregulation of IL-1 Receptor Antagonist in a Mouse Model of Migraine. Brain Sci 2019;9:172. doi:10.3390/brainsci9070172.

[31] Petralia MC, Mazzon E, Fagone P, Falzone L, Bramanti P, Nicoletti F, et al. Retrospective follow-up analysis of the transcriptomic patterns of cytokines, cytokine receptors and chemokines at preconception and during pregnancy, in women with post-partum depression. Exp Ther Med 2019;18:2055–62. doi:10.3892/etm.2019.7774.

[32] Lombardo SD, Mazzon E, Mangano K, Basile MS, Cavalli E, Mammana S, et al. Transcriptomic Analysis Reveals Involvement of the Macrophage Migration Inhibitory Factor Gene Network in Duchenne Muscular Dystrophy. Genes (Basel) 2019;10:939. doi:10.3390/genes10110939.

[33] Lombardo SD, Presti M, Mangano K, Petralia MC, Basile MS, Libra M, et al. Prediction of PD-L1 Expression in Neuroblastoma via Computational Modeling. Brain Sci 2019;9:221. doi:10.3390/brainsci9090221.

[34] Basile MS, Mazzon E, Russo A, Mammana S, Longo A, Bonfiglio V, et al. Differential modulation and prognostic values of immune-escape genes in uveal melanoma. PLoS One 2019;14:e0210276. doi:10.1371/journal.pone.0210276.

[35] Barabási AL, Gulbahce N, Loscalzo J. Network medicine: A network-based approach to human disease. Nat Rev Genet 2011. doi:10.1038/nrg2918.

[36] Guzzi PH, Mercatelli D, Ceraolo C, Giorgi FM. Master Regulator Analysis of the SARS-CoV-2/Human

Interactome. J Clin Med 2020;9:982. doi:10.3390/jcm9040982.

[37] Nelson DE, Ihekwaba AEC, Elliott M, Johnson JR, Gibney CA, Foreman BE, et al. Oscillations in NF-κB signaling control the dynamics of gene expression. Science (80- ) 2004. doi:10.1126/science.1099962.

[38] Xu H, Balakrishnan K, Malaterre J, Beasley M, Yan Y, Essers J, et al. Rad21-cohesin haploinsufficiency impedes DNA repair and enhances gastrointestinal radiosensitivity in mice. PLoS One 2010. doi:10.1371/journal.pone.0012112.

[39] Garrett-Sinha LA, Hou P, Wang D, Grabiner B, Araujo E, Rao S, et al. Spi-1 and Spi-B control the expression of the Grap2 gene in B cells. Gene 2005. doi:10.1016/j.gene.2005.04.009.

[40] Collin M, Dickinson R, Bigley V. Haematopoietic and immune defects associated with GATA2 mutation. Br J Haematol 2015. doi:10.1111/bjh.13317.

[41] Shrimali S, Srivastava S, Varma G, Grinberg A, Pfeifer K, Srivastava M. An ectopic CTCF-dependent transcriptional insulator influences the choice of Vβ gene segments for VDJ recombination at TCRβ locus. Nucleic Acids Res 2012. doi:10.1093/nar/gks556.

[42] Fagone P, Caltabiano R, Russo A, Lupo G, Anfuso CD, Basile MS, et al. Identification of novel chemotherapeutic strategies for metastatic uveal melanoma. Sci Rep 2017;7. doi:10.1038/srep44564.

[43] Cavalli E, Battaglia G, Basile MS, Bruno V, Petralia MC, Lombardo SD, et al. Exploratory Analysis of iPSCS-Derived Neuronal Cells as Predictors of Diagnosis and Treatment of Alzheimer Disease. Brain Sci 2020;10:166. doi:10.3390/brainsci10030166.

[44] Zhou Y, Hou Y, Shen J, Huang Y, Martin W, Cheng F. Network-based drug repurposing for novel coronavirus 2019-nCoV/SARS-CoV-2. Cell Discov 2020. doi:10.1038/s41421-020-0153-3.

[45] Kindrachuk J, Ork B, Hart BJ, Mazur S, Holbrook MR, Frieman MB, et al. Antiviral potential of ERK/MAPK and PI3K/AKT/mTOR signaling modulation for Middle East respiratory syndrome coronavirus infection as identified by temporal kinome analysis. Antimicrob Agents Chemother 2015;59:1088–99. doi:10.1128/AAC.03659-14.

[46] Kuss-Duerkop SK, Wang J, Mena I, White K, Metreveli G, Sakthivel R, et al. Influenza virus differentially activates mTORC1 and mTORC2 signaling to maximize late stage replication. PLoS Pathog 2017. doi:10.1371/journal.ppat.1006635.

[47] Wang CH, Chung FT, Lin SM, Huang SY, Chou CL, Lee KY, et al. Adjuvant treatment with a mammalian target of rapamycin inhibitor, sirolimus, and steroids improves outcomes in patients with severe H1N1 pneumonia and acute respiratory failure. Crit Care Med 2014. doi:10.1097/CCM.0b013e3182a2727d.

[48] Mannick JB, Morris M, Hockey HU, Roma G, Beibel M, Kulmatycki K, et al. TORC1 inhibition enhances immune function and reduces infections in the elderly. Sci Transl Med 2018. doi:10.1126/scitranslmed.aaq1564.

[49] Nicoletti F, Lapenta C, Lamenta C, Donati S, Spada M, Ranazzi A, et al. Inhibition of human

immunodeficiency virus (HIV-1) infection in human peripheral blood leucocytes-SCID reconstituted mice by rapamycin. Clin Exp Immunol 2009;155:28–34. doi:10.1111/j.1365-2249.2008.03780.x.

[50] Nicoletti F, Fagone P, Meroni P, McCubrey J, Bendtzen K. mTOR as a multifunctional therapeutic target in HIV infection. Drug Discov Today 2011;16:715–21.

[51] Donia M, McCubrey JA, Bendtzen K, Nicoletti F. Potential use of rapamycin in HIV infection. Br J Clin Pharmacol 2010;70:784–93. doi:10.1111/j.1365-2125.2010.03735.x.

[52] Heredia A, Le N, Gartenhaus RB, Sausville E, Medina-Moreno S, Zapata JC, et al. Targeting of mTOR catalytic site inhibits multiple steps of the HIV-1 lifecycle and suppresses HIV-1 viremia in humanized mice. Proc Natl Acad Sci U S A 2015;112:9412–7. doi:10.1073/pnas.1511144112.

[53] Latinovic OS, Neal LM, Tagaya Y, Heredia A, Medina-Moreno S, Zapata JC, et al. Suppression of Active HIV-1 Infection in CD34+ Hematopoietic Humanized NSG Mice by a Combination of Combined Antiretroviral Therapy and CCR5 Targeting Drugs. AIDS Res Hum Retroviruses 2019;35:718–28. doi:10.1089/aid.2018.0305.

[54] Stock PG, Barin B, Hatano H, Rogers RL, Roland ME, Lee T-H, et al. Reduction of HIV persistence following transplantation in HIV-infected kidney transplant recipients. Am J Transplant 2014;14:1136–41. doi:10.1111/ajt.12699.

[55] Lai JJ, Lai KP, Zeng W, Chuang KH, Altuwaijri S, Chang C. Androgen receptor influences on body defense system via modulation of innate and adaptive immune systems: Lessons from conditional AR knockout mice. Am J Pathol 2012. doi:10.1016/j.ajpath.2012.07.008.

[56] Bupp MRG, Jorgensen TN. Androgen-induced immunosuppression. Front Immunol 2018. doi:10.3389/fimmu.2018.00794.

[57] Totura AL, Whitmore A, Agnihothram S, Schäfer A, Katze MG, Heise MT, et al. Toll-like receptor 3 signaling via TRIF contributes to a protective innate immune response to severe acute respiratory syndrome coronavirus infection. MBio 2015. doi:10.1128/mBio.00638-15.

[58] Smits SL, van den Brand JMA, de Lang A, Leijten LME, van IJcken WF, van Amerongen G, et al. Distinct Severe Acute Respiratory Syndrome Coronavirus-Induced Acute Lung Injury Pathways in Two Different Nonhuman Primate Species. J Virol 2011. doi:10.1128/jvi.02395-10.

[59] Kovats S. Estrogen receptors regulate innate immune cells and signaling pathways. Cell Immunol 2015. doi:10.1016/j.cellimm.2015.01.018.